\documentclass[epj]{webofc}
\usepackage[utf8]{inputenc}
\usepackage[varg]{txfonts}   
\usepackage{booktabs}
\usepackage{xcolor}
\definecolor{darkred}{rgb}{0.4,0.0,0.0}
\definecolor{darkgreen}{rgb}{0.0,0.4,0.0}
\definecolor{darkblue}{rgb}{0.0,0.0,0.4}
\usepackage[bookmarks,linktocpage,colorlinks,
    linkcolor = darkred,
    urlcolor  = darkblue,
    citecolor = darkgreen]{hyperref}
\usepackage{relsize}
\usepackage{subfigure}
\wocname{EPJ Web of Conferences}
\woctitle{Lattice2017}
\newcommand{\eq}{\begin{equation}}

\newcommand{\eqx}{\end{equation}}
\newcommand{\eqn}{\begin{eqnarray}}
\newcommand{\eqnx}{\end{eqnarray}}
\newcommand{\nn}{\nonumber}
\newcommand{\ra}{\rangle}
\newcommand{\la}{\langle}
\newcommand{\e}{{\rm e\,}}
\begin{document}
%
\selectlanguage{english}
\title{%
Beyond Complex Langevin Equations: \\positive representation of a class of complex measures
}
\author{%
\firstname{Erhard} \lastname{Seiler}\inst{1} \and
\firstname{Jacek}  \lastname{Wosiek}\inst{2}\thanks{Speaker, \email{Jacek.Wosiek@uj.edu.pl} }\fnsep\thanks{Acknowledges financial support by the NCN grant 
UMO-2016/21/B/ST2/01492}
}
\institute{%
Max-Planck-Institut f\"ur Physik (Werner-Heisenberg-Institut), M{\"u}nchen, Germany
\and
M. Smoluchowski Institute of Physics, Jagiellonian University, Cracow, Poland
}
\abstract{%
 A positive representation for a set of complex densities is constructed.
In particular, complex measures on a direct product 
of $U(1)$ groups are studied. 
After identifying general conditions which such representations should satisfy, several concrete 
realizations are proposed. Their utility is  illustrated  in few concrete examples 
representing problems in abelian lattice gauge theories. 
}
\maketitle
\section{The Langevin method - real and complex cases}\label{lanmet}
The Langevin approach is a popular way to replace averaging over given positive probability distribution $\rho(x)=e^{-S(x)}$ by an average over the suitably constructed stochastic process. It's principle is explained by the following chart
\eqn
S(x)\;\;\;\; \stackrel{\dot{x}(\tau)=-\partial_x S +\eta(\tau)}{\longrightarrow}\;\; \;\;\;\;x(\tau)\;\;\; \longrightarrow\;\;\;\;\; P(x,\tau)\;\;\;
\stackrel{\tau\rightarrow\infty}{\longrightarrow} \;\;\; P(x)=e^{-S(x)}\, .\nn
\eqnx
Given a real action $S(x)$, one constructs/generates the stochastic process $x(\tau)$, according to the corresponding Langevin equation.
After the "Langevin time" $\tau$ it's end points are distributed according to $P(x,\tau)$. Under general conditions one can prove that in the infinite
$\tau$ limit this distribution tends to the original $\rho(x)$, which explains the whole idea.

For non-positive, or in general complex, weights $ \rho\equiv e^{-S} $ the approach can still be straightforwardly extended by introducing the complex stochastic process $z(\tau)$
\eqn
S(x)\;\;\;\; \stackrel{\dot{z}(\tau)=-\partial_z S +\eta(\tau)}{\longrightarrow}\;\; \;\;\;\;z(\tau)\;\;\; \longrightarrow\;\;\;\;\; P(x,y,\tau)\, , \;\;\; \;\;\; z \in {\cal} C,\;\;\; \eta \in {\cal R} \, .\nn
\eqnx
This offers a promising possibility to replace an "average" over a complex weight by a statistical average over a positive probability \cite{Pa,Kl} - the task needed badly in many applications of Lattice Field theory:
\eqn
\frac{{\mathlarger{\int}} f(x) \rho(x) dx}{{\mathlarger{\int}} \rho(x) dx } = \frac{{\mathlarger{\int}}{\mathlarger{\int}} f(x+i y) P(x,y) dx dy }{{\mathlarger{\int}}{\mathlarger{\int}} P(x,y) dx dy}. 
\label{B1}
\eqnx
Unfortunately no theorems, involving only conditions on the original weight, exist which relate the large $\tau$ behaviour to the complex action. Consequently the old troubles \cite{AY,HW2}, which had plagued the method, resurfaced again and, 
in spite of substantially better understanding \cite{S4}, compared to the pioneering times , the approach still has serious
difficulties and limitations \cite{Bl,Ph,aarts2017}. For recent review see Ref.\cite{Se}.

\section{Avoiding the trouble - Beyond Complex Langevin}\label{Av}
In view of the above problems, the natural question to ask is wether one can construct the positive distribution $P(x,y)$ from the "matching conditions" (\ref{B1}) alone without any reference to the problematic complex stochastic process at all. The question has been addressed before. In 2002 Weingarten has shown that such a distribution always exists \cite{We}, Salcedo \cite{Sa1,Sa2} constructed  $P$ for the gaussian weights with polynomial modifications. More recently, analytical properties of P in two variables were used to solve the matching conditions, again for a gaussian ( and also a specific quartic ) weight\cite{Wo}. This time the construction was generalized to gaussian quantum mechanical path integrals providing, for the first time, a positive representation for a particle in an external magnetic field directly in the Minkowski time. Until then this text book quantum problem, did not have a positive representation, even after Wick rotation.

In this talk I would like to report on a general prescription how to construct the corresponding positive distributions for complex, weights on a torus $U(1)^N$ \cite{SW}. After  a short presentation of the main principle, some applications ranging from one degree of freedom to small, low dimensional $U(1)$ lattices, will be discussed. An extension for non-compact measures has been already constructed \cite{RW}. See also \cite{WR} for another approach. For generalization for nonabelian integrals see Ref.\cite{Sa5} in this volume.

\section{This work:  periodic weights  }\label{sec:This}

For periodic weights it is natural to rewrite matching conditions (\ref{B1}) in Fourier space (from now on we assume that $\rho$ and $P$ are normalized). Introducing the Fourier components of $\rho(x)$ and partial Fourier transforms of $P(x,y)$
\eqn
\rho(x)=\Sigma_n a_n e^{i n x}\;\;\; and\;\;\; P(x,y)=\Sigma_n P_n(y) e^{i n x} \nn
\eqnx
one rewrites (\ref{B1}) as
\eqn
\mathlarger{\int_{-\pi}^{\pi}} e^{-i n x} \rho(x) dx  = \mathlarger{\int_{-\pi}^{\pi}}\mathlarger{\int_{-\infty}^{\infty}}  e^{-i n ( x + i y)}P(x,y) dx dy . \nn
\eqnx
That is
\eqn
a_n = \mathlarger{\int_{-\infty}^{\infty}} e^{n y} P_n(y) dy \label{an}
\eqnx
It is evident that $P(x,y)$ is not uniquely defined by conditions (\ref{an}). Actually, only one moment of each partial Fourier component
$P_n(y)$ is fixed. This freedom is seen already at the level of the general matching equations (\ref{B1}). It would take at least the full set of two-dimensional moments
\eqn
M_{r,s}=\mathlarger{\int} \mathlarger{\int}  x^r y^s P(x,y) dx dy 
\eqnx
to define uniquely the two-dimensional distribution $P(x,y)$. Instead, as a heritage of the Complex Langevin way of thinking, we imposed in (\ref{B1}) only moments in one,  holomorphic variable \footnote{In this context notice that the starting point of all constructions in \cite{Wo} is the general function $P(z,\bar{z})$ of two variables.}.

Given the above freedom some Ansatz for y-dependence of the partial Fourier components is necessary. We take the simplest one
\eqn
P_n(y)= \lambda_n \delta(y-y_s)+   \mu_n\delta(y+y_s)\, ,  \label{Pn}
\eqnx
and leave the shift $y_s$ as a free parameter. Then the matching equations (\ref{an})  imply
\begin{align}
&\lambda_ n  \e^{ n\cdot y_s} +\mu_n
\e^{- n\cdot y_s}=a_n \,,   
\notag\\
&\lambda_ n \e^{- n\cdot y_s} +\mu_n
\e^{ n\cdot y_s} =a_{-n}^*\,,
\end{align}
with the solutions
\begin{align}
&\lambda_n=\frac{\e^{n y_s}a_n-\e^{-n y_s}a_{-n}^*}{2\sinh(2 n y_s)}  \nn \\
&\mu_n=\frac{\e^{n y_s}a_{-n}^*-\e^{-n y_s }a_n}{2\sinh(2 n y_s)}\,.  \label{sols}
\end{align}
Before we proceed, a few comments are in order.
\begin{itemize}
\item $P(x,y)$ is real by the construction.
\item However it is not positive in general. Positivity can be achieved by the dominance of the lowest mode. This can be realized by choosing large enough $y_s$.
\item Other Ans\"{a}tze are possible, e.g. Gaussian. The gaussian prescription corresponds merely to smearing the point like distributions of the imaginary part $y$. Adjusting a width of these smearing can additionally help to satisfy positivity.
\item Generalization to many variables is straightforward in principle. One basically replaces:  $x, n \longrightarrow \vec{x},\vec{n}$ etc.
\end{itemize}
\section{Examples}
All our examples are built around the problem of the space structure of confining strings. This question attracted attention of lattice community almost since the formulation of lattice QCD \cite{Fuk,HW3,ichie}. The answer is given by the energy density of the colour field in the presence of external $q\bar{q}$ sources. The problem boils down to measuring correlations between an elementary plaquette and a large Wilson loop. Including a Wilson loop in the equivalent, positive measure would dramatically improve results obtained so far.
\subsection{One DOF -- a prototype of a Polyakov line }
We seek for a positive distribution equivalent to the following complex, periodic weight
\eqn
\rho_P(x)=\frac{1}{I_1(\beta)}e^{i x} \exp{\left( 
\beta\cos(x)\right)} = \mathlarger{\sum}_{n=-\infty}^{\infty} \overbrace{\frac{I_{n-1}(\beta)}{I_1(\beta)}}^{a_n} e^{i n x}, \;\;\;<-\pi < x < \pi 
\eqnx
whose Fourier components are well known. Using (\ref{Pn},\ref{sols}) gives immediately
\eqn
P_P(x,y)=P^+(x)\delta(y-y_s)+P^-(x)\delta(y+y_s) \nn
\eqnx
with
\eqn
P^{\pm}(x)=\frac{1}{2} + \sum_{n=1}^{\infty} \cos{(n x)} C^{\pm}_n. \nn
\eqnx 
and
\eqn
C_n^{(\sigma)}=\frac{e^{n \sigma y_s} a_n-e^{-  n \sigma 
y_s}a_{-n}}{\sinh(2 n \sigma 
y_s)},\;\;\;\; 0 < n, \;\;\sigma=\pm 1 .  \nn
\eqnx
Indeed the $P^{\pm}(x)$ are real and positive for large $y_s$. We check explicitly one average
 $<\sin^2(x)>$. Integrating the complex weight gives immediately
\eqn
\int_{-\pi}^{\pi} \frac{dx}{2\pi} \sin^2(x)\rho_P(x) 
=\frac{1}{4 I_1(\beta)}\left(I_1(\beta) - I_3(\beta)\right)\, , \nn
\eqnx
which is readily reproduced by the positive density integral.
\eqn
\int_{-\pi}^{\pi} \frac{dx}{2\pi} dy \sin^2(x+i y)P(x,y)= \int \frac{dx}{2\pi} \sin^2{(x+i 
y_s)}\left\{\frac{1}{2}+\sum_{n=1}^{\infty} \cos{(n x)} C^+_n\right\}\nn\\
+\int \frac{dx}{2\pi} \sin^2{(x-i 
y_s)}\left\{\frac{1}{2}+\sum_{n=1}^{\infty} \cos{(n x)} C^-_n\right\}=\nn\\
\frac{1}{2} - \frac{1}{4}\cosh{(2y_s)} \left(C^+_2+C^-_2\right)=
\frac{1}{4I_1(\beta)}\left(I_1(\beta)-I_3(\beta)\right)\, . \nn
\eqnx 

\subsection{Four DOF with gauge invariance -- Wilson loop}

Our second example contains four link angles $x_i$ with the miniscule gauge invariance: $x_i \rightarrow x_i+\alpha$.
The unnormalized complex density reads
\eqn
\rho_P(x_1,x_2,x_3,x_4)=
e^{i(x_1+x_2-x_3-x_4)}\exp{\left(\beta\cos(x_1+x_2-x_3-x_4)\right)} = 
\sum_{\vec{n}} a_{\vec{n}} e^{i \vec{n}\cdot\vec{x}} \nn 
\eqnx
where the phase factor represents now a small Wilson loop. Again the Fourier components are simple

\eqn
a_{\vec{n}}= \sum_m I_{m-1} 
\delta_{m,n_1}\delta_{m,n_2}\delta_{m,-n_3}\delta_{m,-n_4}\, . \nn
\eqnx
For the corresponding positive distribution $P_P(\vec{x},\vec{y})$ we take now
\eqn
P_P(\vec{x},\vec{y})=\delta(\vec{y}-\vec{y}_s) P^+(x)+\delta(\vec{y}+\vec{y}_s) P^-(x),\;\;\;\vec{y}_s=y_s(1,1,-1,-1), \nn
\eqnx
which essentally reproduces the previous example
\eqn
P^{\sigma}(\vec{x})=\frac{ I_1}{2}+\sum_{m, m \ne 0} 
 \frac{e^{ 4 m y_s } I_{m-\sigma}}{\sinh{( 8 m y_s )}}\cos{\left( m 
(x_1+x_2-x_3-x_4\right)}\, , \nn
\eqnx 
up to a simple rescaling of the shift parameter.
\newpage
\subsection{ Tiny 2D abelian lattice  }
In the last example we put 2 Polyakov lines on a 2x2 U(1) lattice and construct equivalent positive distribution.
Links and plaquettes are labeled as in Fig. 1. 

\begin{figure}[thb]
\centering
  \sidecaption
  \includegraphics[width=4cm]{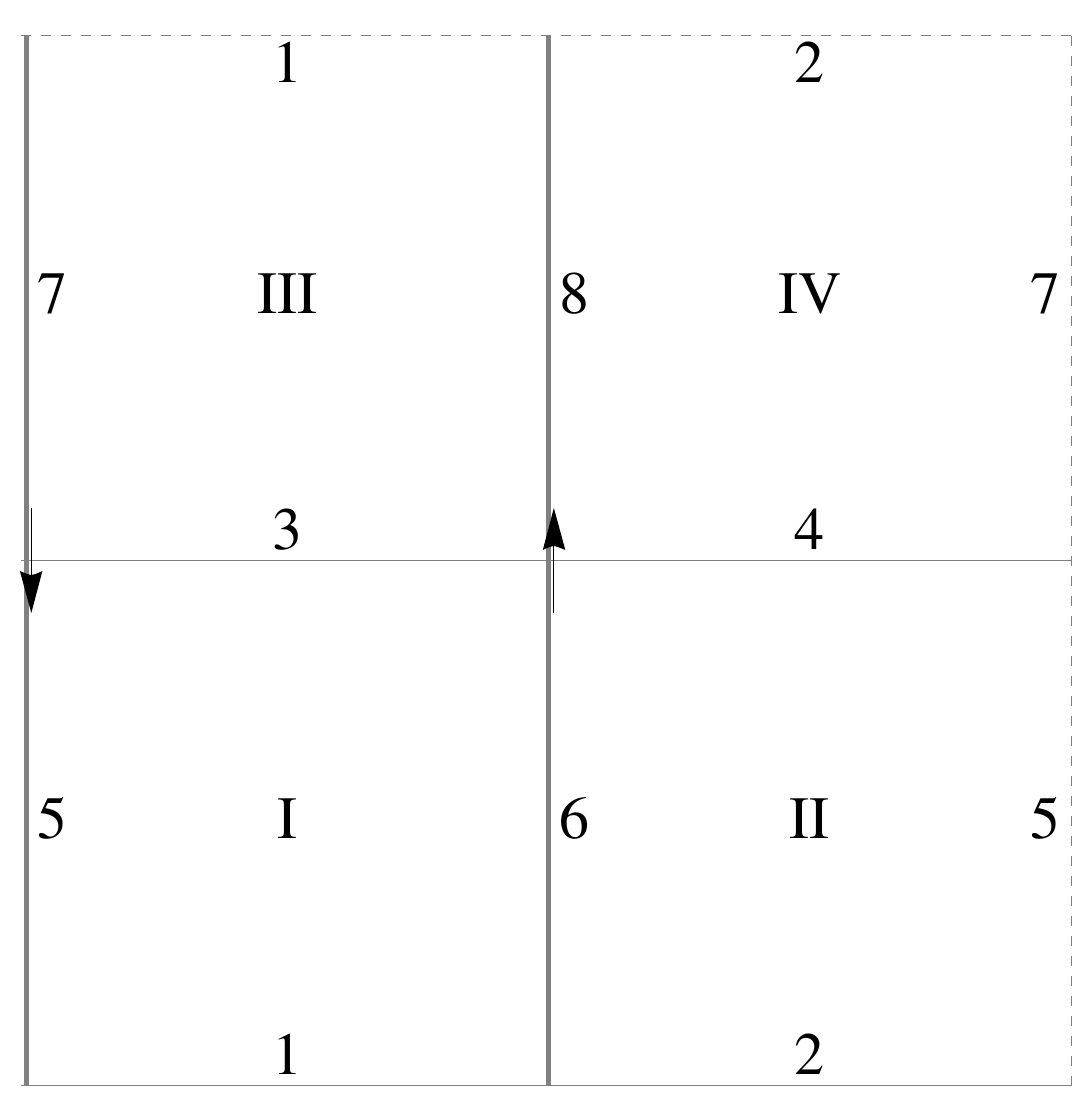}
   \caption{A tiny periodic lattice with two Polyakov lines.}
  \label{fig-2}
\end{figure}
\noindent The complex density reads (for simplicity we denote link angles by their indices $\theta_i \rightarrow i$).
\eqn
\rho(\vec{\theta})=&B(3+8-1-7)B(4+7-2-8)\nn\\
                    &B(1+6-3-5)B(2+5-4-6)\label{B2}\\
                    &U(-5-7)U(6+8)\nn
 \eqnx
 where 
\eqn        
 \theta_i \rightarrow i,\;\;\;\;B(\phi)=\exp(\beta \cos(\phi)),\;\;\;\;U(\phi)=\exp(i \phi),\;\;\;
\eqnx
There are only three independent variables and in this simple example. One can take them to be any three plaquette angles. We choose
 $( \phi_{I},\phi_{II},\phi_{III} ) \rightarrow ( \phi_{1},\phi_{2},\phi_{3} ) $. Then 
\eqn
\rho(\vec{\phi})= B(\phi_1) B(\phi_2)B(\phi_3) B(\phi_1+\phi_2+\phi_3) 
U(\phi_1) U(\phi_3) \nn
\eqnx
Fourier components of $\rho$ are again simple
\eqn
a_{\vec{n}}= \sum_m I_{m} I_{m-n_2} I_{m-n_1+1} I_{m-n_3+1}\, ,\;\;\;\; \vec{n}=(n_1,n_2,n_3) \nn
\eqnx
and one can readily construct the corresponding positive density $(\vec \phi = \vec x+ i \vec y)$.
\eqn
P(\vec x,\vec y)&=&\frac{a_{\vec{0}}}{2}\delta(\vec{y}-\vec{y}_s)+\frac{a_{\vec{0}}}{2}\delta(\vec{y}+\vec{y}_s)\\
\lefteqn{+\sum_{\vec{n}\ne\vec{0} } e^{i \vec{n}\cdot\vec{x}}\left\{\frac{e^{\vec{n}\cdot\vec{y}_s}a_{\vec{n}}-e^{-\vec{n}\cdot\vec{y}_s}a_{-\vec{n}}}{2\sinh{(2 \vec{n}\cdot\vec{y}_s)}}\delta(\vec{y}-\vec{y}_s)+\frac{e^{\vec{n}\cdot\vec{y}_s}a_{-\vec{n}}-e^{-\vec{n}\cdot\vec{y}_s}a_{\vec{n}}}{2\sinh{(2 \vec{n}\cdot\vec{y}_s)}}\delta(\vec{y}+\vec{y}_s)\right\} }\nn
\eqnx
To avoid singularities introduced by zeroes of $\vec{n}\cdot\vec{y}_s$, we took $\vec{y}_s=y_s (1,\sqrt{2},\sqrt{3})$. It is a simple matter to check that indeed $P$ reproduces all moments of the complex weight
\eqn
\la \left(e^{i \phi_1}\right)^{r_1}  \left(e^{i \phi_2}\right)^{r_2}  \left(e^{i \phi_3}\right)^{r_3} \ra_{\rho(\vec{\phi})} = \la \left(e^{i (x_1+i y_1)}\right)^{r_1} \left(e^{i (x_2+i y_2)}\right)^{r_2}  \left(e^{i (x_3+i y_3)}\right)^{r_3} \ra_{P(\vec{x},\vec{y})} 
\eqnx
as it should.

It is also instructive to examine directly the effect of complex phases, cf. Fig.2.

The influence of the complex phases is dramatic. The effective distribution differs essentially from the original Boltzmann density. This confirms explicitly the common sense expectations, that the regions of field space contributing substantially to both averages are very different. For the first time one is able to replace the averaging over the complex Polyakov or Wilson loops by a standard, statistical average with respect to a positive distribution.  

\begin{figure}[h]
\begin{center}
\includegraphics[width=8cm]{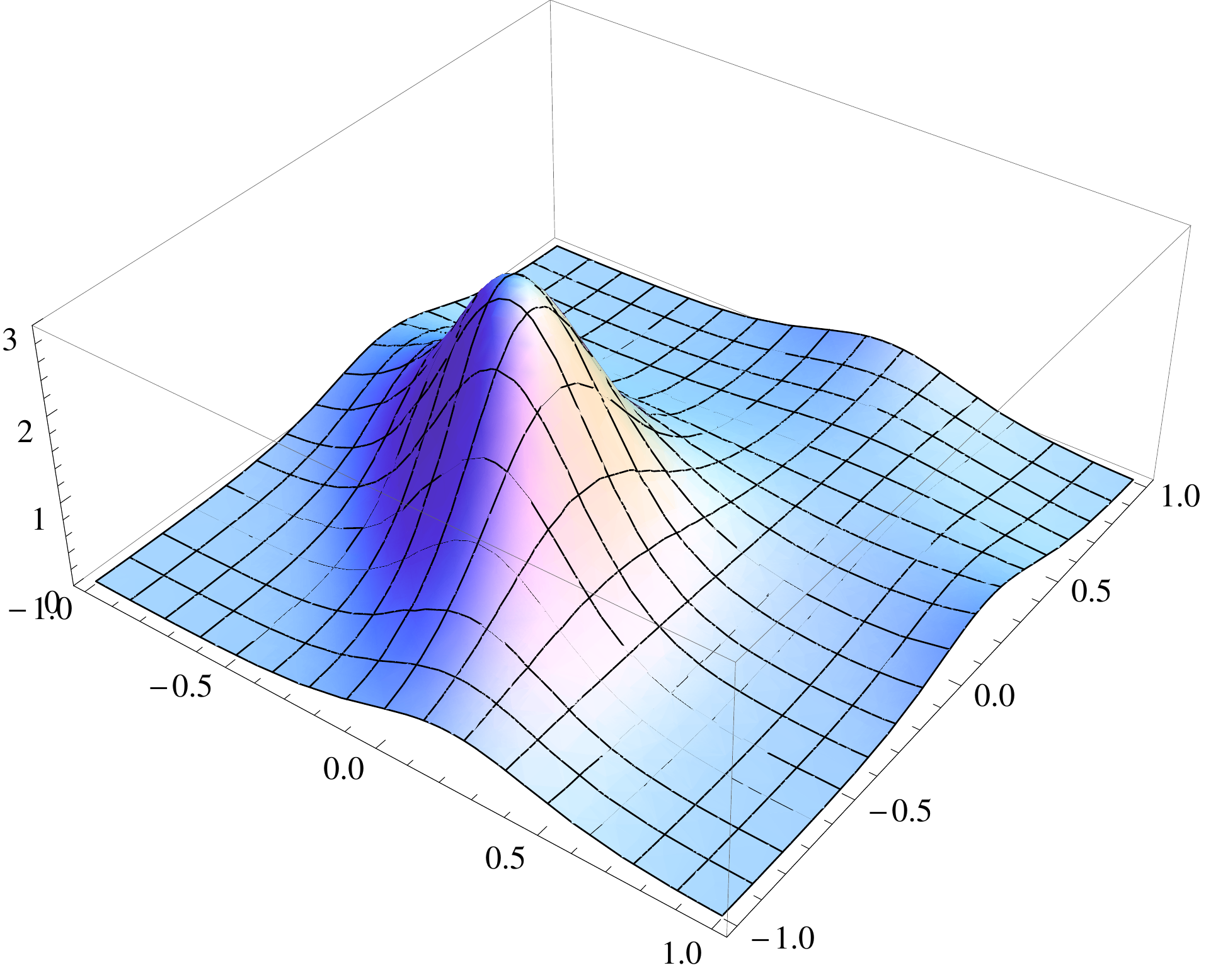}  \vspace*{1cm} \newline
\includegraphics[width=8cm]{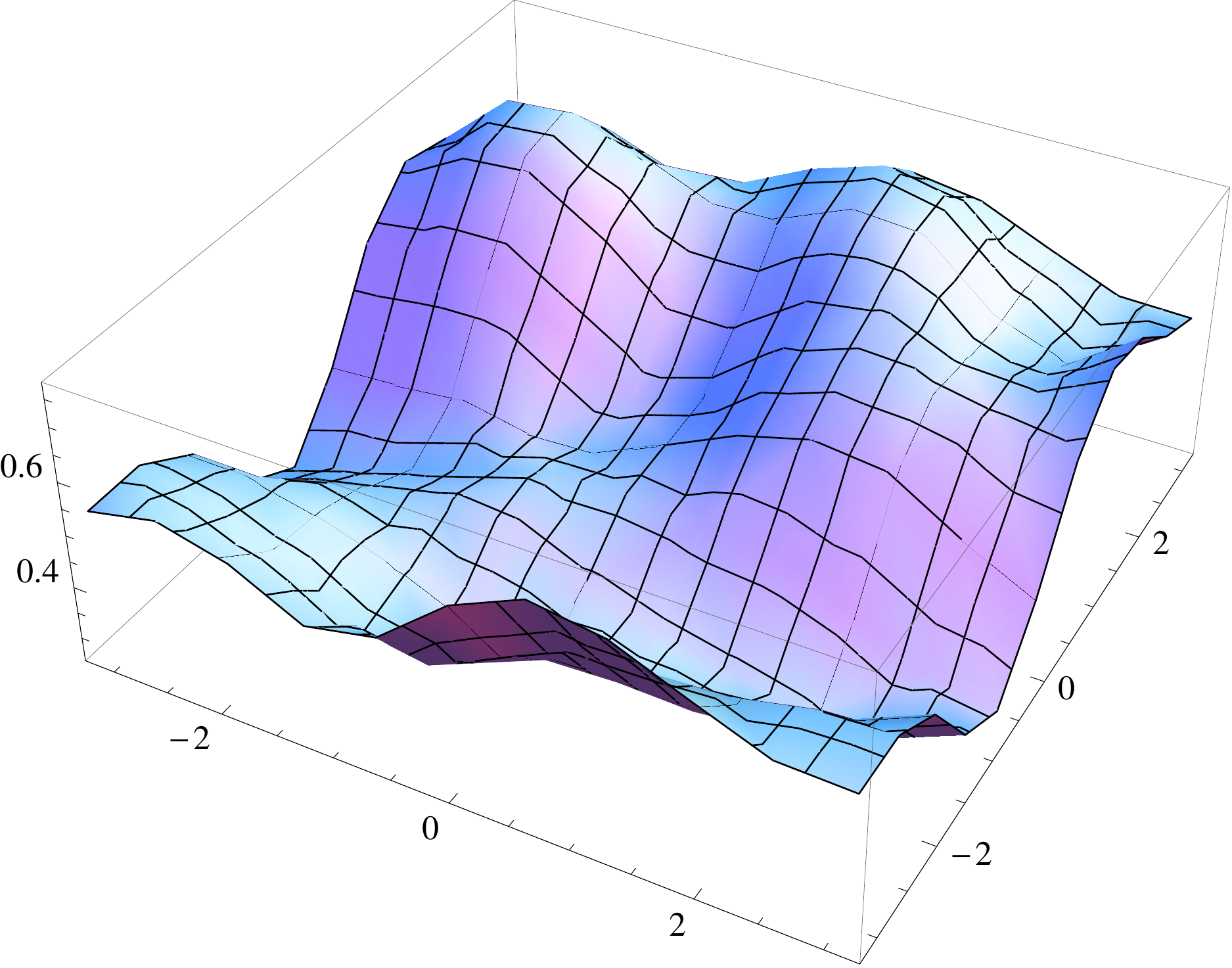}
\end{center}
\vskip-4mm \caption{The effect of including Polyakov lines into the positive distribution. Upper panel: original positive Boltzmann factor, as in Eq.(\ref{B2}) without U's. Lower: $P^+(\vec x)$ component of the positive distribution equivalent to the full complex weight, Eq.(\ref{B2}) \label{fig:f2}. Profiles depend on three variables. Two of them extend over the whole Brillouin zone, while the third one is fixed at $0.6 \pi$,  $\beta=1$ and the shift parameter $y_s=8$, which is sufficient to ensure positivity. Results for $P^-$ are similar.}
\end{figure}

The second good news is that the variation of $P^+$ is substantial.This means that the dominance
of the first, i.e. (0, 0, 0) mode, required in the proof of positivity, does not preclude
the importance of other modes. Consequently  the effective positive distribution reveals an interesting structure.

\section{Generalizations, summary and an outlook}
\begin{figure}[h]
\begin{center}
\includegraphics[width=5cm]{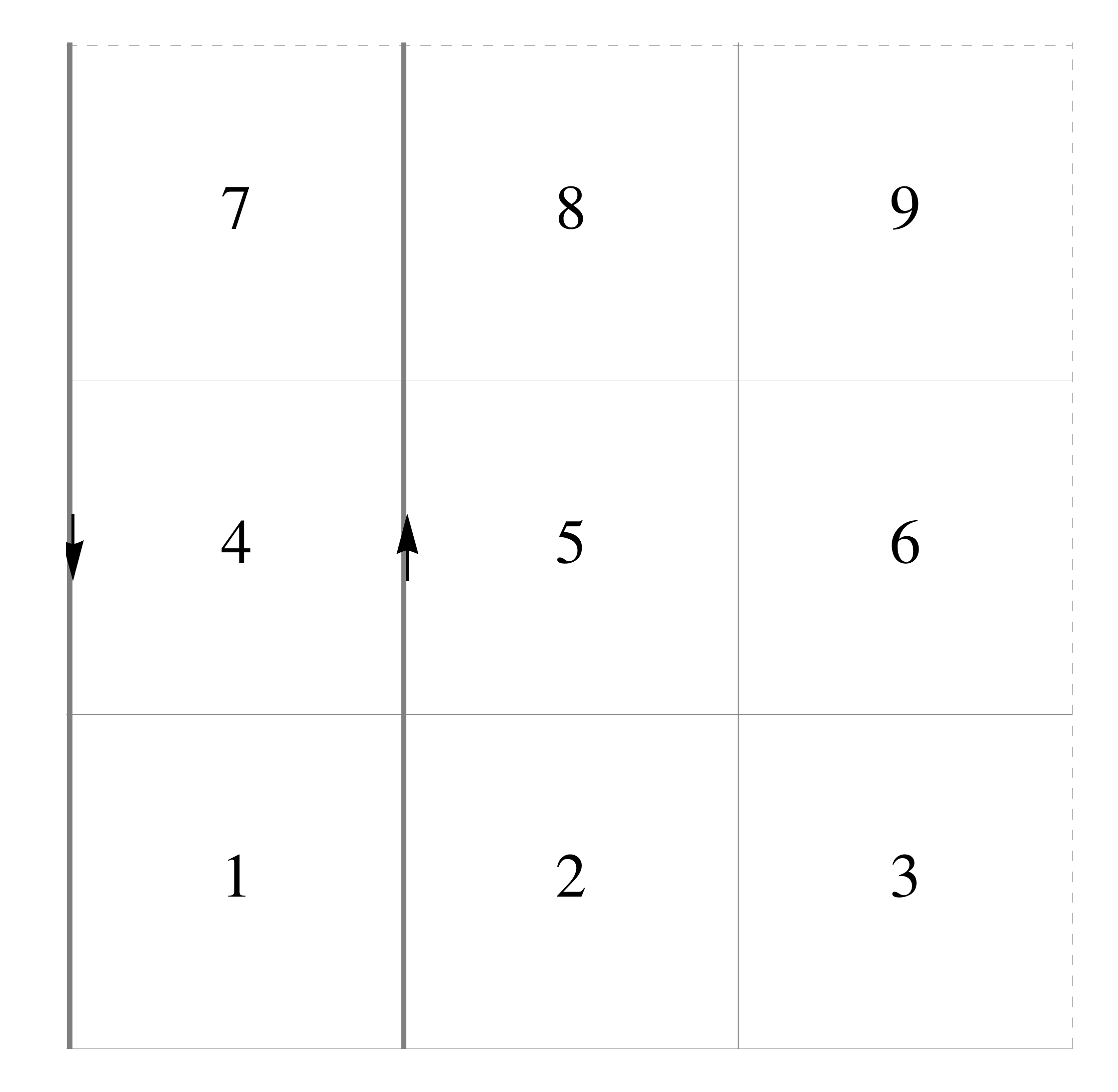} \hspace*{3cm}
\includegraphics[width=5cm]{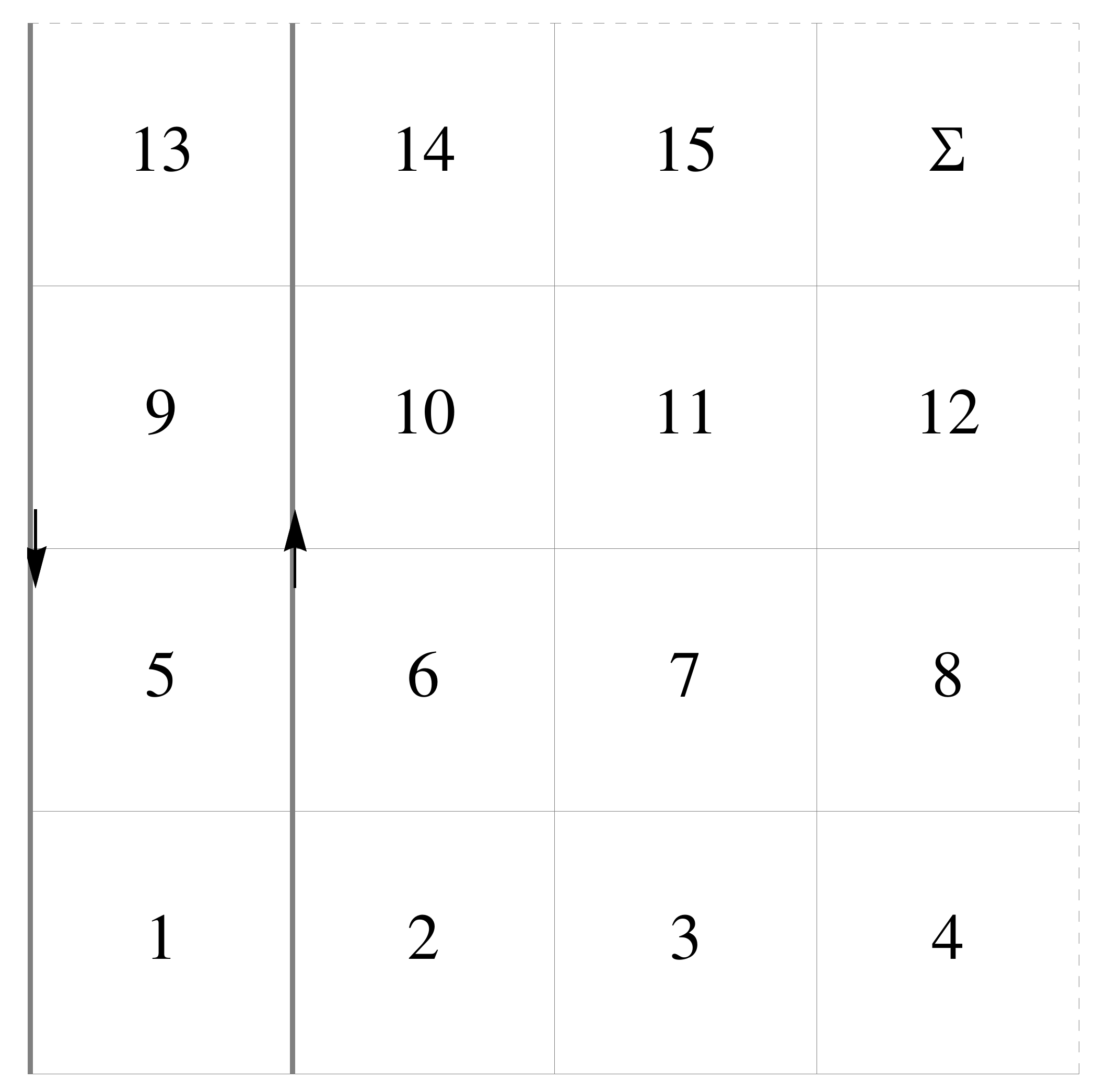}
\end{center}
\vskip-4mm \caption{Towards larger abelian lattices. One plaquette, denoted $\Sigma$ in the second panel depends linearly on the rest for periodic lattices.}\label{f3}
\end{figure}

Extension of this example to larger lattices is straightforward, c.f. Fig.\ref{f3}. 

For example, the complex weight for 3x3 lattice reads

\eqn
\rho(1...8)=B(1)B(2)...B(8)B(1+2+3+...+8)U(1+4+7)\, , \label{rho3} \nn
\eqnx
and the corresponding Fourier components are
\eqn
a_{\vec{n}}= \sum_m  
(I_{m-n_7+1} I_{m-n_8}I_{m})\,
(I_{m-n_4+1} I_{m-n_5}I_{m-n_6})\,
(I_{m-n_1+1} I_{m-n_2}I_{m-n_3})\, . \nn
\eqnx
Therefore, at least for these simple models, the present construction provides positive solutions, even for many variables.
This generalizes readily to arbitrary, abelian (2D) lattices with the complex density in the form (ignoring, or choosing suitable boundary conditions).
\eqn
\rho(\vec{\phi})=\left(\prod_{exterior\; of\; W} B(\phi_{ext})\right) 
\left( \prod_{interior\; of\; W} B(\phi_{int}) U(\phi_{int}) \right)\, . \label{any}
\eqnx
Of course, some practical questions remain. For example inversion of the multidimensional Fourier transform will be expensive for more variables. Using the Fast Fourier  
Transforms might alleviate the problem. Also for the separable systems, like in (\ref{any}), this is not an issue, since the complexification procedure can be carried variable by variable and resulting positive distributions would also factorize. In the general case, however, even with local nearest neighbour interactions in a complex weight, positive distributions do not have to be local.

At the same time an interesting possibility appears. As explained earlier, matching conditions  do not determine the positive distribution uniquely. There is a large freedom in constructing $P(x,y)$. Conceivably it can be used to satisfy additional requirements imposed on $P$, possibly locality.

Summarizing, one can avoid poorly convergent stochastic processes by constructing a positive distribution $P(x,y)$ directly from the matching conditions (\ref{B1}).
At first sight this task looks rather formidable, but after closer scrutiny it turns out to be underdetermined. One way to solve the problem is to rewrite (\ref{B1}) for the Fourier modes, satisfy relations with the aid of some simple ansatz, and go back to the original, configuration space. Positivity is assured by arranging the dominance of the lowest, constant mode. We have shown that this construction works not only as a matter of principle, but also in a range of concrete examples, which cover simple lattice systems, also with few variables. 

With larger lattices our solution, even though correct in principle, gives in general non-local distributions. With tremendous developments of  numerical algorithms, and matching growth of computing power, non-locality is not as severe obstacle, as it was in early days of Lattice Field Theory. Still, of course, it is important to look for improvements. One possibility was already proposed in Ref.\cite{Sa3} and should be studied further. Another interesting option is to exploit the inherent ambiguity of the formulation. Possibly the large freedom, exposed many times in this talk, may be used to construct more local, positive distributions.


\end{document}